\documentclass[%
 reprint,
 twocolumn,
 showpacs,
 showkeys,
 preprintnumbers,
 amsmath,amssymb,
 aps,
  pra,
  longbibliography,
  floatfix,
 ]{revtex4-1}

\usepackage[greek,american]{babel}

\usepackage{tikz}
\usepackage[breaklinks=true,colorlinks=true,anchorcolor=blue,citecolor=blue,filecolor=blue,menucolor=blue,pagecolor=blue,urlcolor=blue,linkcolor=blue]{hyperref}
\usepackage{graphicx}% Include figure files
\usepackage{url}

\usepackage{xcolor}
\usepackage{colortbl}

\usepackage{eurosym}

\begin{document}
%\selectlanguage{american}

\title{Induction and physical theory formation as well as universal computation by machine learning}

%\cdmtcsauthor{Karl Svozil}
%\cdmtcsaffiliation{Vienna University of Technology}
%\cdmtcstrnumber{407}
%\cdmtcsdate{September 2011}
%\coverpage

\author{Alexander Svozil}
\affiliation{Theory and Applications of Algorithms, Faculty of Computer Science, University of Vienna\\
W\"hringer Stra\ss e 29, A-1090 Vienna, Austria}
\email{alexander.svozil@univie.ac.at}
\homepage[\\]{https://taa.cs.univie.ac.at/team/person/64683/}

\author{Karl Svozil}
\affiliation{Institute for Theoretical Physics, Vienna
    University of Technology, Wiedner Hauptstra\ss e 8-10/136, A-1040
    Vienna, Austria}
\email{svozil@tuwien.ac.at} \homepage[]{http://tph.tuwien.ac.at/~svozil}

%\preprint{CDMTCS preprint nr. 407/2011}

\begin{abstract}
Machine learning presents a general, systematic framework for the generation of formal theoretical models for physical description and prediction. Tentatively standard linear modeling techniques are reviewed; followed by a brief discussion of generalizations to deep forward networks for approximating nonlinear phenomena and universal computers.
\end{abstract}

\maketitle

%% https://www.google.at/webhp?sourceid=chrome-instant&ion=1&espv=2&ie=UTF-8#q=machine%20learning%20physics

\section{Algorithmic induction}

There appear to be at least two approaches to induction.
The first route is by intuition and ingenuity. It has been successfully pursued by geniuses and gifted individuals.
A typical representative of this approach to knowledge is Ramanujan
who seemed to have attributed his revelations to a Hindu Goddess~\cite{Kanigel-1991}.
In western thought, this is often more secularly referred to as Platonism.
G\"odel seemed to have held the opinion that our minds have access to truth, which can be discovered through
personal insights -- maybe even beyond the bounds of universal computability --
in particular, the idea that  minds are no (Turing) machines~\cite[p.~216]{kreisel-80}.
As successful these narratives may have been, they remain anecdotal and cannot be generalized.

When it comes to {\it ad hoc} revelations of individuals, there may also be psychological issues.
These have been described by Freud~\cite{Freud-1912}, pointing to the dangers
caused by {``temptations to project,
what  [the analyst]  in dull self-perception recognizes as the peculiarities of his own personality,
as generally valid theory into science.''}
A similar warning comes from Jaynes'  ``Mind Projection Fallacy''~\cite{jaynes-89,jaynes-90}, pointing out that
{\em ``we are all under an ego-driven temptation to project our private
thoughts out onto the real world, by supposing that the creations of one's own imagination are real
properties of Nature, or that one's own ignorance signifies some kind of indecision on the part of
Nature.''}

A second, computational, approach could be conceived in the spirit of Turing~\cite{turing1948intelligent}.
In this line of thought, it is possible to obtain knowledge about a system by mechanical, algorithmic procedures;
such as a deterministic agent
{\em ``provided with paper, pencil and rubber, and subject to strict discipline
[carrying out a set of rules of procedure written down]''}~\cite[p.~34]{Turing-Intelligent_Machinery}.
Recently one of the more promising methods to algorithmic induction has been machine learning~\cite{Goodfellow-et-al-2016-Book}
which will be pursued here.
There is even a form of collective intuition -- the so-called
{\em swarm intelligence}~\cite{bonabeau-Swarm-1999,Eberhart-Swarm-2001,Zelinka-2018} --
that can solve many logistic problems effectively.

Two {\it caveats} should be stated upfront.
First, the general induction problem---in recursion theoretic terms, the rule inference problem---is unsolvable with respect and relative to
universal computational capacities~\cite{go-67,blum75blum,angluin:83,ad-91}.
So, in certain (even constructible) situations machine learning, like all other algorithmic induction strategies, provably fails.
But that does not exclude heuristic methods of induction, such as machine learning applied to physical phenomena.
Second, theoretical constructions cannot be expected to faithfully represent the ``laws underlying'' a phenomenology.
As Lakatos~\cite{lakatosch} points out, the progressiveness and degeneracy of research programs are transient; and
without a recognizable coherent conceptual convergence.
Therefore, the ``explanations'' and theoretical models generated by machine learning  present
knowledge and explanations which cannot claim to have any ontologic (only epistemic) relevance --
they are means relative to the respective methods employed.

Whereas machine learning has already been applied to very specific problems in high energy~\cite{2014-Higgs-ml}
and solid-state physics~\cite{Arsenault-2015},
we would like to propose it as an extremely general method of theory formation and induction.

We are  applying the obtained machine representations to predict a simulacrum -- we are, in particular, interested in universal Turing computation; at least until
a finite amount of computational space and time~\cite{gandy1}.

\section{Linear Models of inertial motion}

In what follows a {\em linear regression} model~\cite[Sect.~5.1.4]{Goodfellow-et-al-2016-Book} will be used.
Thereby we shall, if not stated otherwise explicitly, closely follow the notation of Mermin's book on {\em Quantum Computer Science} \cite{mermin-07}.

Suppose, for the sake of demonstration, a one-dimensional physical system of a particle in inertial motion.
Suppose further that it has been (approximately) measured already at $n \ge 2$ positions
$x_1, \ldots , x_n$
at times
$t_1, \ldots , t_n$,
respectively.
The goal is to find a general algorithm which predicts its location at an arbitrary time $\tau$.

In what follows the respective positions and times are (not necessarily successively) arranged as $n$-tuples; that is,
as a finite ordered list of elements, and interpreted as $(n \times 1)$-matrices
\begin{equation}
\begin{split}
\vert {\bf x} \rangle \equiv  \begin{pmatrix} x_{i_1}, \ldots , x_{i_n} \end{pmatrix}^\intercal \\
\vert {\bf t} \rangle \equiv  \begin{pmatrix} t_{i_1}, \ldots , t_{i_n} \end{pmatrix}^\intercal;
\end{split}
\end{equation}
whereby
the superscript $^\intercal$ indicates transposition,
and
$i_1,\ldots , i_n$ are arbitrary permutations of $1,\ldots ,n$.
That is, it is not necessary to order the events temporally; actually, they can ``run backward''
or be randomly arranged~\cite{permutationcity}.

A linear regression {\it Ansatz} is to find a linear model for the prediction of some unknown
observable, given some anecdotal instances of its performance.
More formally, let
$y$ be an arbitrary observable which depends
on $n$ parameters $x_1, \ldots , x_n$  by linear means; that is, by
\begin{equation}
y = \sum_{i=1}^n x_i r_i = \langle  {\bf x}\vert {\bf r} \rangle,
\label{2016-ml-ansatz-lr}
\end{equation}
where $\langle {\bf x} \vert = (\vert {\bf x} \rangle )^\intercal$ is the transpose
of the vector $\vert {\bf x} \rangle$,  the tuple
\begin{equation}
\vert {\bf r} \rangle = \begin{pmatrix} r_1, \ldots , r_n \end{pmatrix}^\intercal
\label{2016-ml-ansatz-vectorweights}
\end{equation}
contains the unknown weights of the approximation --
the ``theory,'' if you like --
and $\langle  {\bf a}\vert {\bf b} \rangle = \sum_i a_ib_i$ stands for the Euclidean scalar product of the tuples interpreted
as (dual) vectors in $n$-dimensional (dual) vector space $\mathbb{R}^n$.

Given are $m$ known instances of (\ref{2016-ml-ansatz-lr}); that is, suppose $m$ pairs
$\begin{pmatrix}z_j, \vert {\bf x}_j \rangle \end{pmatrix}$ are known.
These data can be bundled into an $m$-tuple
\begin{equation}
\vert {\bf z} \rangle \equiv \begin{pmatrix} z_{j_1}, \ldots , z_{j_m} \end{pmatrix}^\intercal,
\end{equation}
and an $(m \times n)$-matrix
\begin{equation}
\textsf{\textbf{X}} \equiv
\begin{pmatrix}
x_{{j_1}{i_1}} & \ldots & x_{{j_1}{i_n}}\\
\vdots & \vdots & \vdots \\
x_{{j_m}{i_1}} & \ldots & x_{{j_m}{i_n}}
\end{pmatrix}
\end{equation}
where $j_1,\ldots , j_m$ are arbitrary permutations of $1,\ldots ,m$,
and the matrix rows are just the vectors
$\vert {\bf x}_{j_k} \rangle \equiv \begin{pmatrix} x_{{j_k}{i_1}} & \ldots , x_{{j_k}{i_n}} \end{pmatrix}^\intercal$.

The task is to compute a ``good'' estimate of $\vert {\bf r} \rangle$;
that is, an estimate of $\vert {\bf r} \rangle$
which allows an ``optimal'' computation of the prediction $y$.

Suppose that a good way to measure the performance
of the prediction from some particular definite but unknown $\vert {\bf r} \rangle $
with respect to the $m$ given data
$\begin{pmatrix}z_j, \vert {\bf x}_j \rangle \end{pmatrix}$
is by the mean squared error (MSE)
\begin{equation}
\begin{split}
\text{MSE}
=
\frac{1}{m}
\left\|
\vert {\bf y} \rangle - \vert {\bf z} \rangle
\right\|^2
=
\frac{1}{m}
\left\|
\textsf{\textbf{X}} \vert {\bf r} \rangle
 - \vert {\bf z}   \rangle
\right\|^2
\\
=
\frac{1}{m}
\left(
\textsf{\textbf{X}} \vert {\bf r} \rangle
 - \vert {\bf z}   \rangle
\right)^\intercal
\left(
\textsf{\textbf{X}} \vert {\bf r} \rangle
 - \vert {\bf z}   \rangle
\right)
\\
=
\frac{1}{m}
\left(
\langle {\bf r} \vert \textsf{\textbf{X}}^\intercal
- \langle {\bf z} \vert
\right)
\left(
\textsf{\textbf{X}} \vert {\bf r} \rangle
- \vert {\bf z}   \rangle
\right)
\\
=
\frac{1}{m} \left(
\langle {\bf r} \vert \textsf{\textbf{X}}^\intercal \textsf{\textbf{X}} \vert {\bf r} \rangle
- \langle {\bf z} \vert  \textsf{\textbf{X}} \vert {\bf r} \rangle
- \langle {\bf r} \vert \textsf{\textbf{X}}^\intercal   \vert {\bf z}   \rangle
+ \langle {\bf z} \vert    {\bf z}   \rangle
\right)
\\
=
\frac{1}{m} \left[
\langle {\bf r} \vert \textsf{\textbf{X}}^\intercal \textsf{\textbf{X}} \vert {\bf r} \rangle
- \langle {\bf z} \vert \left( \langle  {\bf r} \vert \textsf{\textbf{X}}^\intercal\right)^\intercal
- \langle {\bf r} \vert \textsf{\textbf{X}}^\intercal   \vert {\bf z}   \rangle
+ \langle {\bf z} \vert    {\bf z}   \rangle
\right]
\\
=
\frac{1}{m} \left\{
\langle {\bf r} \vert \textsf{\textbf{X}}^\intercal \textsf{\textbf{X}} \vert {\bf r} \rangle
- \left[  \left( \langle  {\bf r} \vert \textsf{\textbf{X}}^\intercal \right)^\intercal  \right]^\intercal \vert {\bf z} \rangle
\right .\\ \left.
- \langle {\bf r} \vert \textsf{\textbf{X}}^\intercal   \vert {\bf z}   \rangle
+ \langle {\bf z} \vert    {\bf z}   \rangle
\right\}
\\
=
\frac{1}{m} \left(
\langle {\bf r} \vert \textsf{\textbf{X}}^\intercal \textsf{\textbf{X}} \vert {\bf r} \rangle
- 2 \langle {\bf r} \vert \textsf{\textbf{X}}^\intercal   \vert {\bf z}   \rangle
+ \langle {\bf z} \vert    {\bf z}   \rangle
\right)
.
\end{split}
\label{2016-ml-MSE}
\end{equation}

In order to minimize the mean squared error (\ref{2016-ml-MSE}) with respect to variations of $\vert {\bf r} \rangle$
one obtains a condition for ``the linear theory'' $\vert {\bf y} \rangle$
by setting its derivatives (its gradient) to zero; that is
\begin{equation}
\partial_{ \vert {\bf r} \rangle } \text{MSE}
=
{\bf 0}.
\label{2016-ml-MSE-min}
\end{equation}
A lengthy but straightforward computation yields
\begin{equation}
\begin{split}
\frac{\partial }{\partial r_i}
\left(
r_j  \textsf{\textbf{X}}^\intercal_{jk} \textsf{\textbf{X}}_{kl} r_l -2 r_j \textsf{\textbf{X}}^\intercal_{jk}z_k + z_jz_j
\right)
\\
=
\delta_{ij}  \textsf{\textbf{X}}^\intercal_{jk} \textsf{\textbf{X}}_{kl} r_l
+
r_j \textsf{\textbf{X}}^\intercal_{jk} \textsf{\textbf{X}}_{kl}  \delta_{il}
-2 \delta_{ij} \textsf{\textbf{X}}^\intercal_{jk}z_k
\\
=
\textsf{\textbf{X}}^\intercal_{ik} \textsf{\textbf{X}}_{kl} r_l
+
r_j \textsf{\textbf{X}}^\intercal_{jk} \textsf{\textbf{X}}_{ki}
-2 \textsf{\textbf{X}}^\intercal_{ik}z_k
\\
=
\textsf{\textbf{X}}^\intercal_{ik} \textsf{\textbf{X}}_{kl} r_l
+
\textsf{\textbf{X}}^\intercal_{ik} \textsf{\textbf{X}}_{kj} r_j
-2 \textsf{\textbf{X}}^\intercal_{ik}z_k
\\
=
2 \textsf{\textbf{X}}^\intercal_{ik} \textsf{\textbf{X}}_{kj} r_j
-2 \textsf{\textbf{X}}^\intercal_{ik} z_k
\\
\equiv
2\left( \textsf{\textbf{X}}^\intercal \textsf{\textbf{X}} \vert {\bf r} \rangle - \textsf{\textbf{X}}^\intercal   \vert {\bf z} \rangle
\right)
=0
\end{split}
\label{2016-ml-MSE-min-res-der}
\end{equation}
and finally, upon multiplication with
$ \left( \textsf{\textbf{X}}^\intercal \textsf{\textbf{X}}  \right)^{-1}$ from the left,
\begin{equation}
\vert {\bf r} \rangle
=  \left( \textsf{\textbf{X}}^\intercal \textsf{\textbf{X}} \right)^{-1}
\textsf{\textbf{X}}^\intercal \vert {\bf z} \rangle
.
\label{2016-ml-MSE-min-res}
\end{equation}
A short plausibility check for $n=m=1$ yields the linear dependency
$\vert {\bf z} \rangle  =  \textsf{\textbf{X}} \vert {\bf r} \rangle$.

Coming back to the one-dimensional physical system of a particle in inertial motion,
we could characterize inertial motion in machine learning and linear regression terms
by the requirement that the Ansatz (\ref{2016-ml-ansatz-lr}) is ``good'' in the sense that
predictions can be made to a ``sufficient degree'' (a term which is arbitrary, subjective and thus conventional);
that is, within a pre-defined error.

If the particle does not pass through the origin, it might be necessary to augment Eq.~(\ref{2016-ml-ansatz-lr})
with an affine term $d$, which can be absorbed into
$\vert {\bf x}' \rangle = \begin{pmatrix} x_1', \ldots , x_n', x_{n+1}' \end{pmatrix}^\intercal
= \begin{pmatrix} x_1, \ldots , x_n, 1 \end{pmatrix}^\intercal $
and
$\vert {\bf r}' \rangle = \begin{pmatrix} r_1', \ldots , r_n', r_{n+1}' \end{pmatrix}^\intercal
= \begin{pmatrix} r_1, \ldots , r_n, d \end{pmatrix}^\intercal $
such that
\begin{equation}
y = \sum_{i=1}^n x_i r_i = \langle  {\bf x}\vert {\bf r} \rangle + d = \sum_{i=1}^{n+1} x'_i r'_i
.
\label{2016-ml-ansatz-lr-affine}
\end{equation}

\section{Nonlinear Models of noninertial motion}

The linear Ansatz (\ref{2016-ml-ansatz-lr}) fails for noninertial motion.
One possible way to cope with nonlinearities would be to introduce extra dimensions
corresponding to nonlinear terms, such as $x^l$ for  $2\le l \le d<\infty$;
with the consequence that the dimensionality of the parameter space increases.

To cope with nonlinear phenomena,
{\em deep forward networks}
have been used in machine learning~\cite[Chapt.~6]{Goodfellow-et-al-2016-Book}.
This strategy of deep learning
invokes intermediate {\em hidden theoretical layers} of description which communicate with each other.
For the sake of an example, suppose there are two functions
$g$ and $h$, connected in a chain by functional substitution, such that
$f(x)=h(g(x))$. The length of the chain is identified with the {\em depth} of the model -- in this case two.
$g$ is the first layer of the model.
The final layer -- in this case $h$ -- is called the {\em output layer.}

Unfortunately, the linear regression Ansatz (\ref{2016-ml-ansatz-lr-affine})
for $g$ and $h$ would effectively be linear again.
Therefore, to model a nonlinear phenomenology,
a nonlinear Ansatz for at least one layer has to be implemented.
Such networks
are capable of approximating any Borel measurable function (and its derivative, even if it a generalized function)
from one finite-dimensional space to another~\cite{Hornik1989359,hornik:Hornik+Stinchcombe+White:1990,Hornik1991251}.

\section{Simulation of universal Turing machines by deep forward networks}

It could be suspected that, even though for all practical purposes~\cite{bell-a}
the methods and techniques
discussed so far are ``good,''
the task of finding ``better and better'' approximations or even total correspondence might,
at least in some cases, turn out to be ``difficult'' if not outrightly impossible.
Because suppose, for the sake of a {\it reductio ad absurdum} -- more precisely, a reduction to the rule inference and halting problems --
that such learning of the exact behavior would be computable.
Such a suspicion might be tempting, considering the approximate ``solution''
of the general rule inference problem~\cite{go-67,blum75blum,angluin:83,ad-91}
suggested earlier in the limit of infinite precision.
Alas, any such unbounded computation, as long as it needs to be finite,
would run into the problem that no computable rate of convergence can be given
-- very much like the Busy Beaver function~\cite{chaitin-bb} or Chaitin's halting probability $\Omega$~\cite{2002-glimpseofran,calude-dinneen06}.
Formally, any such claim can be reduced to the rule inference and the halting problem of universal computers.
The situation is not dissimilar to series solutions of the $n$-body problem~\cite{Wang91},
which may converge ``very slowly'' (indeed, intractably slow in numerical work~\cite{Diacu96}); but if the system encodes a universal computer,
they cannot converge in general due to reduction to the halting problem~\cite{svozil-2007-cestial}.

However, one could (courageously) ``invert'' or transform these objections and deficiencies into virtues~\cite{Nietzsche-JGB,Nietzsche-GM}
and argue that,
for all practical purposes and in many relevant instantiations, machine learning, and, in particular,
deep forward networks, may turn out to be effective in the simulation of universal computers
such as a universal Turing machine.
Thereby the criterium is not to obtain an exact correspondence; rather the result of the simulation should
be ``good enough'' to justify its use.

For the sake of an example, think of the typical physical estimate of a quantity in terms of
{\em orders of magnitude}: very often, the applicability of a suggestion or technique
does not depend on the exact observable value it generates but rather on the order of its impact.
This is not dissimilar to certain quantum advantages: for instance, the Deutsch algorithm identifies
the parity or (non)constancy of a binary function of a single bit without identifying the actual function~\cite{mermin-07};
thereby rendering a partitioning of the set of all such individual functions.

How could one imagine training a deep forward network to simulate a universal computer?
Fairly simple: let it access the input-output behavior of actual ``exact'' devices with
von Neuman architecture. After ``lots of'' training, the network should be able to emulate
the performance of this architecture within ``reasonable precision.''
That is, it won't be able to give the exact value of, say, an algebraic operation like
$n+m$ for ``large'' (for physical realizability)
numbers $n$ and $m$. But it might be able to output some estimate which
is ``close to'' (for the applicability) the exact result.

One could also say that, in this scheme, the deep forward network acts as an oracle with respect to the universal computer.
And although the training of it may take some time because of the sheer training volume,
as well as the conceivable computational complexity of the individual input-output functions involved,
eventually, the trained network is not bound by these restrictions and can reach an (approximate) result quite fast;
such that the computation time it takes for any simulation is uniformly bounded from above.
In particular, the halting problem can be said to be ``for all practical purposes (FAPP) solvable'' by such oracles;
but, of course, no guarantee of validity or total precision can be given.
Inadvertedly one cannot exclude instances in which the deep forward network predicts halting whereas the universal
computer it simulates does not halt, and {\it vice versa}.

One may also ask: where exactly is the physical resource rendering the computational
capacity of such ``universal'' deep forward networks located?
There has to be some formal symbolic encoding in terms of physical components or entities
making the simulation feasible and effective.
One rather straightforward way to answer this is in terms of the {\em connections} or {\em correlations}
among the nodes involved: if they are modeled as continuous formal entities such as real or complex numbers
then the capacity of even a finite such configuration to store information is unbounded.

In more pragmatic, practical terms
the nonexact but effective simulation of general (universal) computations may also present
a way to circumvent the stall in Moore's law which can be observed already for a couple of years.
Currently, because of physical restrictions on circuit and switch designs,
most improvements in performance are due to parallelization rather than miniaturization.
Eventually, the switching time of electronic devices is restricted
by fundamental limits from below on resistance; in particular the von Klitzing constant $R_k$.

\section{Discussion}

One objection for applying machine learning algorithms to physical theory creation
or simulation of universal computers might be that the resulting representations
lack any sort of ``meaning;'' that is, that these representations amount to pure syntax devoid of any conceptual
semantics. But if conceptual semantics is omitted,
there can be no true ``understanding'' of the ``physics behind'' the phenomena,
or the computational processes yielding those estimates.

One may counter this criticism by noticing that, first,
underlying such objections is the premise that something {\em can} actually be discovered or revealed.
This realistic ontology is by far nontrivial and is heavily debated~\cite{berkeley,stace1,Goldschmidt2017-idealism}.
If, for example, the phenomena emerge from primordial chaos, such as
in Greek mythology and cosmology, \textgreek{q'aoc},
then any ``meaning'' one might present and ``discover'' ultimately remains a (pragmatic) narrative or a
mathematical abstraction such as Ramsey theory at best:
for any data, there cannot be no correlations -- regardless of the origin or type of empirical data,
there always has to be some, maybe spurious~\cite{Calude2016,svozil-2018-was}, regularity or coincidences or properties.
How can it be excluded that the laws of physics are nothing but yet undiscovered consequences of Ramsey theory?

Second, as has already been pointed out, historic evidence seems to suggest that
successive physical conceptual models (say, of gravity) are not continuously evolving;
but that they are disruptive and dissimilar~\cite{kuhn,lakatosch}:
they lack conceptual convergence.
One may even go so far as to suggest that,
in any case,
theories are (more or less~\cite{feyerabend}) successful belief systems;
very much like Greek mythology~\cite{Veyne-Greek-Mythology}.

Third, also the present perception of the quantum mechanical formalism includes, among other inclinations,
the position that no interpretation is necessary~\cite{fuchs-peres}; that indeed, interpretation
is even dangerous and detrimental for the researcher~\cite[p.~129]{feynman-law};
or that, at the very least, there are no issues with respect to
interpretation~\cite{Englert2013}.

Nevertheless, it might be quite amusing to study toy universes capable of universal computation, such as Conway's {\em game of life},
{\it via} intrinsic, embedded, machine learning algorithms.
It could not be excluded that these kinds of algorithmic agents ``come up'' with the ``right rules;''
that is those rules which define the toy mini-universe.
It can be expected that if a machine learning
algorithm performs excellently on particular problems then
it necessarily has a degraded performance on the
set of all remaining problems~\cite{Wolpert:1996:LPD:1362127.1362128,Wolpert:1997:NFL:2221336.2221408}.

In any case, the ways how physical theories are created and invented by human individuals is not dissimilar from machine learning.
And machine learning might become of great practical utility for the simulation of (universal) computations.

\acknowledgments{
This work was supported in part by the European Union, Research Executive Agency (REA),
Marie Curie FP7-PEOPLE-2010-IRSES-269151-RANPHYS grant.

Responsibility for the information and views expressed in this article lies entirely with the authors.

The authors declare no conflict of interest.

}

%\bibliography{svozil}

\begin{thebibliography}{46}%
\makeatletter
\providecommand \@ifxundefined [1]{%
 \@ifx{#1\undefined}
}%
\providecommand \@ifnum [1]{%
 \ifnum #1\expandafter \@firstoftwo
 \else \expandafter \@secondoftwo
 \fi
}%
\providecommand \@ifx [1]{%
 \ifx #1\expandafter \@firstoftwo
 \else \expandafter \@secondoftwo
 \fi
}%
\providecommand \natexlab [1]{#1}%
\providecommand \enquote  [1]{``#1''}%
\providecommand \bibnamefont  [1]{#1}%
\providecommand \bibfnamefont [1]{#1}%
\providecommand \citenamefont [1]{#1}%
\providecommand \href@noop [0]{\@secondoftwo}%
\providecommand \href [0]{\begingroup \@sanitize@url \@href}%
\providecommand \@href[1]{\@@startlink{#1}\@@href}%
\providecommand \@@href[1]{\endgroup#1\@@endlink}%
\providecommand \@sanitize@url [0]{\catcode `\\12\catcode `\$12\catcode
  `\&12\catcode `\#12\catcode `\^12\catcode `\_12\catcode `\%12\relax}%
\providecommand \@@startlink[1]{}%
\providecommand \@@endlink[0]{}%
\providecommand \url  [0]{\begingroup\@sanitize@url \@url }%
\providecommand \@url [1]{\endgroup\@href {#1}{\urlprefix }}%
\providecommand \urlprefix  [0]{URL }%
\providecommand \Eprint [0]{\href }%
\providecommand \doibase [0]{http://dx.doi.org/}%
\providecommand \selectlanguage [0]{\@gobble}%
\providecommand \bibinfo  [0]{\@secondoftwo}%
\providecommand \bibfield  [0]{\@secondoftwo}%
\providecommand \translation [1]{[#1]}%
\providecommand \BibitemOpen [0]{}%
\providecommand \bibitemStop [0]{}%
\providecommand \bibitemNoStop [0]{.\EOS\space}%
\providecommand \EOS [0]{\spacefactor3000\relax}%
\providecommand \BibitemShut  [1]{\csname bibitem#1\endcsname}%
\let\auto@bib@innerbib\@empty
%</preamble>
\bibitem [{\citenamefont {Kanigel}(1991)}]{Kanigel-1991}%
  \BibitemOpen
  \bibfield  {author} {\bibinfo {author} {\bibfnamefont {Robert}\ \bibnamefont
  {Kanigel}},\ }\href@noop {} {\emph {\bibinfo {title} {The Man Who Knew
  Infinity: A Life of the Genius {R}amanujan}}},\ \bibinfo {edition} {5th}\
  ed.\ (\bibinfo  {publisher} {Washington Square Press},\ \bibinfo {year}
  {1991})\BibitemShut {NoStop}%
\bibitem [{\citenamefont {Kreisel}(1980)}]{kreisel-80}%
  \BibitemOpen
  \bibfield  {author} {\bibinfo {author} {\bibfnamefont {Georg}\ \bibnamefont
  {Kreisel}},\ }\bibfield  {title} {\enquote {\bibinfo {title} {{K}urt
  {G}\"odel. 28 {A}pril 1906-14 {J}anuary 1978},}\ }\href {\doibase
  10.1098/rsbm.1980.0005} {\bibfield  {journal} {\bibinfo  {journal}
  {Biographical memoirs of Fellows of the Royal Society}\ }\textbf {\bibinfo
  {volume} {26}},\ \bibinfo {pages} {148--224} (\bibinfo {year} {1980})},\
  \bibinfo {note} {corrections {\it Ibid.} {\bf 27}, 697; {\it ibid.} {\bf 28},
  718}\BibitemShut {NoStop}%
\bibitem [{\citenamefont {Freud}(1912, 1999)}]{Freud-1912}%
  \BibitemOpen
  \bibfield  {author} {\bibinfo {author} {\bibfnamefont {Sigmund}\ \bibnamefont
  {Freud}},\ }\bibfield  {title} {\enquote {\bibinfo {title}
  {{R}atschl{\"{a}}ge f{\"{u}}r den {A}rzt bei der psychoanalytischen
  {B}ehandlung},}\ }in\ \href
  {http://gutenberg.spiegel.de/buch/kleine-schriften-ii-7122/15} {\emph
  {\bibinfo {booktitle} {{G}esammelte {W}erke. {C}hronologisch geordnet.
  {A}chter {B}and. {W}erke aus den {J}ahren 1909--1913}}},\ \bibinfo {editor}
  {edited by\ \bibinfo {editor} {\bibfnamefont {Anna}\ \bibnamefont {Freud}},
  \bibinfo {editor} {\bibfnamefont {E.}~\bibnamefont {Bibring}}, \bibinfo
  {editor} {\bibfnamefont {W.}~\bibnamefont {Hoffer}}, \bibinfo {editor}
  {\bibfnamefont {E.}~\bibnamefont {Kris}}, \ and\ \bibinfo {editor}
  {\bibfnamefont {O.}~\bibnamefont {Isakower}}}\ (\bibinfo  {publisher}
  {Fischer},\ \bibinfo {address} {Frankfurt am Main},\ \bibinfo {year} {1912,
  1999})\ pp.\ \bibinfo {pages} {376--387}\BibitemShut {NoStop}%
\bibitem [{\citenamefont {Jaynes}(1989)}]{jaynes-89}%
  \BibitemOpen
  \bibfield  {author} {\bibinfo {author} {\bibfnamefont {Edwin~Thompson}\
  \bibnamefont {Jaynes}},\ }\bibfield  {title} {\enquote {\bibinfo {title}
  {Clearing up mysteries - the original goal},}\ }in\ \href
  {http://bayes.wustl.edu/etj/articles/cmystery.pdf} {\emph {\bibinfo
  {booktitle} {Maximum-Entropy and Bayesian Methods: Proceedings of the 8th
  Maximum Entropy Workshop, held on August 1-5, 1988, in St. John's College,
  Cambridge, England}}},\ \bibinfo {editor} {edited by\ \bibinfo {editor}
  {\bibfnamefont {John}\ \bibnamefont {Skilling}}}\ (\bibinfo  {publisher}
  {Kluwer},\ \bibinfo {address} {Dordrecht},\ \bibinfo {year} {1989})\ pp.\
  \bibinfo {pages} {1--28}\BibitemShut {NoStop}%
\bibitem [{\citenamefont {Jaynes}(1990)}]{jaynes-90}%
  \BibitemOpen
  \bibfield  {author} {\bibinfo {author} {\bibfnamefont {Edwin~Thompson}\
  \bibnamefont {Jaynes}},\ }\bibfield  {title} {\enquote {\bibinfo {title}
  {Probability in quantum theory},}\ }in\ \href
  {http://bayes.wustl.edu/etj/articles/prob.in.qm.pdf} {\emph {\bibinfo
  {booktitle} {Complexity, Entropy, and the Physics of Information: Proceedings
  of the 1988 Workshop on Complexity, Entropy, and the Physics of Information,
  held May - June, 1989, in Santa Fe, New Mexico}}},\ \bibinfo {editor} {edited
  by\ \bibinfo {editor} {\bibfnamefont {Wojciech~Hubert}\ \bibnamefont
  {Zurek}}}\ (\bibinfo  {publisher} {Addison-Wesley},\ \bibinfo {address}
  {Reading, MA},\ \bibinfo {year} {1990})\ pp.\ \bibinfo {pages}
  {381--404}\BibitemShut {NoStop}%
\bibitem [{\citenamefont {Turing}(1996)}]{turing1948intelligent}%
  \BibitemOpen
  \bibfield  {author} {\bibinfo {author} {\bibfnamefont {Alan~M.}\ \bibnamefont
  {Turing}},\ }\bibfield  {title} {\enquote {\bibinfo {title} {Intelligent
  machinery, a heretical theory},}\ }\href {\doibase 10.1093/philmat/4.3.256}
  {\bibfield  {journal} {\bibinfo  {journal} {Philosophia Mathematica}\
  }\textbf {\bibinfo {volume} {4}},\ \bibinfo {pages} {256--260} (\bibinfo
  {year} {1996})}\BibitemShut {NoStop}%
\bibitem [{\citenamefont {Turing}(1968)}]{Turing-Intelligent_Machinery}%
  \BibitemOpen
  \bibfield  {author} {\bibinfo {author} {\bibfnamefont {Alan~M.}\ \bibnamefont
  {Turing}},\ }\bibfield  {title} {\enquote {\bibinfo {title} {Intelligent
  machinery},}\ }in\ \href@noop {} {\emph {\bibinfo {booktitle} {{C}ybernetics.
  {K}ey Papers}}},\ \bibinfo {editor} {edited by\ \bibinfo {editor}
  {\bibfnamefont {C.~R.}\ \bibnamefont {Evans}}\ and\ \bibinfo {editor}
  {\bibfnamefont {A.~D.~J.}\ \bibnamefont {Robertson}}}\ (\bibinfo  {publisher}
  {Butterworths},\ \bibinfo {address} {London},\ \bibinfo {year} {1968})\ pp.\
  \bibinfo {pages} {27--52}\BibitemShut {NoStop}%
\bibitem [{\citenamefont {Goodfellow}\ \emph {et~al.}(2016)\citenamefont
  {Goodfellow}, \citenamefont {Bengio},\ and\ \citenamefont
  {Courville}}]{Goodfellow-et-al-2016-Book}%
  \BibitemOpen
  \bibfield  {author} {\bibinfo {author} {\bibfnamefont {Ian}\ \bibnamefont
  {Goodfellow}}, \bibinfo {author} {\bibfnamefont {Yoshua}\ \bibnamefont
  {Bengio}}, \ and\ \bibinfo {author} {\bibfnamefont {Aaron}\ \bibnamefont
  {Courville}},\ }\href {https://mitpress.mit.edu/books/deep-learning} {\emph
  {\bibinfo {title} {Deep Learning}}}\ (\bibinfo {year} {2016})\BibitemShut
  {NoStop}%
\bibitem [{\citenamefont {Bonabeau}\ \emph {et~al.}(1999)\citenamefont
  {Bonabeau}, \citenamefont {Dorigo},\ and\ \citenamefont
  {Theraulaz}}]{bonabeau-Swarm-1999}%
  \BibitemOpen
  \bibfield  {author} {\bibinfo {author} {\bibfnamefont {Eric}\ \bibnamefont
  {Bonabeau}}, \bibinfo {author} {\bibfnamefont {Marco}\ \bibnamefont
  {Dorigo}}, \ and\ \bibinfo {author} {\bibfnamefont {Guy}\ \bibnamefont
  {Theraulaz}},\ }\href
  {https://global.oup.com/academic/product/swarm-intelligence-9780195131598}
  {\emph {\bibinfo {title} {Swarm Intelligence: {F}rom Natural to Artificial
  Systems}}},\ Santa Fe Institute Studies on the Sciences of Complexity\
  (\bibinfo  {publisher} {Oxford University Press},\ \bibinfo {address} {New
  York, NY},\ \bibinfo {year} {1999})\BibitemShut {NoStop}%
\bibitem [{\citenamefont {Kennedy}\ \emph {et~al.}(2001)\citenamefont
  {Kennedy}, \citenamefont {Eberhart},\ and\ \citenamefont
  {Shi}}]{Eberhart-Swarm-2001}%
  \BibitemOpen
  \bibfield  {author} {\bibinfo {author} {\bibfnamefont {James}\ \bibnamefont
  {Kennedy}}, \bibinfo {author} {\bibfnamefont {Russell~C.}\ \bibnamefont
  {Eberhart}}, \ and\ \bibinfo {author} {\bibfnamefont {Yuhui}\ \bibnamefont
  {Shi}},\ }\href {\doibase 10.1016/B978-1-55860-595-4.X5000-1} {\emph
  {\bibinfo {title} {Swarm Intelligence}}},\ The Morgan Kaufmann Series in
  Evolutionary Computation\ (\bibinfo  {publisher} {Morgan Kaufmann, Academic
  Press, Elsevier},\ \bibinfo {address} {San Francisco, San Diego, CA},\
  \bibinfo {year} {2001})\BibitemShut {NoStop}%
\bibitem [{\citenamefont {Zelinka}\ and\ \citenamefont
  {Chen}(2018)}]{Zelinka-2018}%
  \BibitemOpen
  \bibfield  {author} {\bibinfo {author} {\bibfnamefont {Ivan}\ \bibnamefont
  {Zelinka}}\ and\ \bibinfo {author} {\bibfnamefont {Guanrong}\ \bibnamefont
  {Chen}},\ }\href {\doibase 10.1007/978-3-662-55663-4} {\emph {\bibinfo
  {title} {Evolutionary Algorithms, Swarm Dynamics and Complex Networks.
  {M}ethodology, Perspectives and Implementation}}},\ \bibinfo {series}
  {Emergence, Complexity and Computation}, Vol.~\bibinfo {volume} {26}\
  (\bibinfo  {publisher} {Springer-Verlag},\ \bibinfo {address} {Berlin,
  Heidelberg},\ \bibinfo {year} {2018})\BibitemShut {NoStop}%
\bibitem [{\citenamefont {Gold}(1967)}]{go-67}%
  \BibitemOpen
  \bibfield  {author} {\bibinfo {author} {\bibfnamefont {Mark~E.}\ \bibnamefont
  {Gold}},\ }\bibfield  {title} {\enquote {\bibinfo {title} {Language
  identification in the limit},}\ }\href {\doibase
  10.1016/S0019-9958(67)91165-5} {\bibfield  {journal} {\bibinfo  {journal}
  {Information and Control}\ }\textbf {\bibinfo {volume} {10}},\ \bibinfo
  {pages} {447--474} (\bibinfo {year} {1967})}\BibitemShut {NoStop}%
\bibitem [{\citenamefont {Blum}\ and\ \citenamefont {Blum}(1975)}]{blum75blum}%
  \BibitemOpen
  \bibfield  {author} {\bibinfo {author} {\bibfnamefont {Lenore}\ \bibnamefont
  {Blum}}\ and\ \bibinfo {author} {\bibfnamefont {Manuel}\ \bibnamefont
  {Blum}},\ }\bibfield  {title} {\enquote {\bibinfo {title} {Toward a
  mathematical theory of inductive inference},}\ }\href {\doibase
  10.1016/S0019-9958(75)90261-2} {\bibfield  {journal} {\bibinfo  {journal}
  {Information and Control}\ }\textbf {\bibinfo {volume} {28}},\ \bibinfo
  {pages} {125--155} (\bibinfo {year} {1975})}\BibitemShut {NoStop}%
\bibitem [{\citenamefont {Angluin}\ and\ \citenamefont
  {Smith}(1983)}]{angluin:83}%
  \BibitemOpen
  \bibfield  {author} {\bibinfo {author} {\bibfnamefont {Dana}\ \bibnamefont
  {Angluin}}\ and\ \bibinfo {author} {\bibfnamefont {Carl~H.}\ \bibnamefont
  {Smith}},\ }\bibfield  {title} {\enquote {\bibinfo {title} {Inductive
  inference: Theory and methods},}\ }\href {\doibase 10.1145/356914.356918}
  {\bibfield  {journal} {\bibinfo  {journal} {ACM Computing Surveys}\ }\textbf
  {\bibinfo {volume} {15}},\ \bibinfo {pages} {237--269} (\bibinfo {year}
  {1983})}\BibitemShut {NoStop}%
\bibitem [{\citenamefont {Adleman}\ and\ \citenamefont {Blum}(1991)}]{ad-91}%
  \BibitemOpen
  \bibfield  {author} {\bibinfo {author} {\bibfnamefont {Leonard~M.}\
  \bibnamefont {Adleman}}\ and\ \bibinfo {author} {\bibfnamefont {Manuel}\
  \bibnamefont {Blum}},\ }\bibfield  {title} {\enquote {\bibinfo {title}
  {Inductive inference and unsolvability},}\ }\href {\doibase 10.2307/2275058}
  {\bibfield  {journal} {\bibinfo  {journal} {The Journal of Symbolic Logic}\
  }\textbf {\bibinfo {volume} {56}},\ \bibinfo {pages} {891--900} (\bibinfo
  {year} {1991})}\BibitemShut {NoStop}%
\bibitem [{\citenamefont {Lakatos}(1978)}]{lakatosch}%
  \BibitemOpen
  \bibfield  {author} {\bibinfo {author} {\bibfnamefont {Imre}\ \bibnamefont
  {Lakatos}},\ }\href@noop {} {\emph {\bibinfo {title} {Philosophical Papers.
  1. {T}he Methodology of Scientific Research Programmes}}}\ (\bibinfo
  {publisher} {Cambridge University Press},\ \bibinfo {address} {Cambridge},\
  \bibinfo {year} {1978})\BibitemShut {NoStop}%
\bibitem [{\citenamefont {Team}(2014)}]{2014-Higgs-ml}%
  \BibitemOpen
  \bibfield  {author} {\bibinfo {author} {\bibfnamefont {Kaggle}\ \bibnamefont
  {Team}},\ }\href {https://higgsml.lal.in2p3.fr/} {\enquote {\bibinfo {title}
  {The {H}iggs{ML} challenge: When high energy physics meets machine
  learning},}\ } (\bibinfo {year} {2014}),\ \bibinfo {note} {may-September
  2014, accessed August 31, 2016}\BibitemShut {NoStop}%
\bibitem [{\citenamefont {Arsenault}\ \emph {et~al.}(2015)\citenamefont
  {Arsenault}, \citenamefont {von Lilienfeld},\ and\ \citenamefont
  {Millis}}]{Arsenault-2015}%
  \BibitemOpen
  \bibfield  {author} {\bibinfo {author} {\bibfnamefont {Louis-Francois}\
  \bibnamefont {Arsenault}}, \bibinfo {author} {\bibfnamefont {O.~Anatole}\
  \bibnamefont {von Lilienfeld}}, \ and\ \bibinfo {author} {\bibfnamefont
  {Andrew~J.}\ \bibnamefont {Millis}},\ }\href
  {https://arxiv.org/abs/1506.08858} {\enquote {\bibinfo {title} {Machine
  learning for many-body physics: efficient solution of dynamical mean-field
  theory},}\ } (\bibinfo {year} {2015}),\ \Eprint
  {http://arxiv.org/abs/arXiv:1506.08858} {arXiv:1506.08858} \BibitemShut
  {NoStop}%
\bibitem [{\citenamefont {Gandy}(1982)}]{gandy1}%
  \BibitemOpen
  \bibfield  {author} {\bibinfo {author} {\bibfnamefont {Robin~O.}\
  \bibnamefont {Gandy}},\ }\bibfield  {title} {\enquote {\bibinfo {title}
  {Limitations to mathematical knowledge},}\ }in\ \href
  {https://www.elsevier.com/books/logic-colloquium-80/van-dalen/978-0-444-86465-9}
  {\emph {\bibinfo {booktitle} {Logic colloquium '80}}},\ \bibinfo {editor}
  {edited by\ \bibinfo {editor} {\bibfnamefont {D.}~\bibnamefont {van Dalen}},
  \bibinfo {editor} {\bibfnamefont {D.}~\bibnamefont {Lascar}}, \ and\ \bibinfo
  {editor} {\bibfnamefont {J.}~\bibnamefont {Smiley}}}\ (\bibinfo  {publisher}
  {North Holland},\ \bibinfo {address} {Amsterdam},\ \bibinfo {year} {1982})\
  pp.\ \bibinfo {pages} {129--146},\ \bibinfo {note} {papers intended for the
  {E}uropean Summer Meeting of the Association for Symbolic Logic}\BibitemShut
  {NoStop}%
\bibitem [{\citenamefont {Mermin}(2007)}]{mermin-07}%
  \BibitemOpen
  \bibfield  {author} {\bibinfo {author} {\bibfnamefont {David~N.}\
  \bibnamefont {Mermin}},\ }\href {\doibase 10.1017/CBO9780511813870} {\emph
  {\bibinfo {title} {Quantum Computer Science}}}\ (\bibinfo  {publisher}
  {Cambridge University Press},\ \bibinfo {address} {Cambridge},\ \bibinfo
  {year} {2007})\BibitemShut {NoStop}%
\bibitem [{\citenamefont {Egan}(1994)}]{permutationcity}%
  \BibitemOpen
  \bibfield  {author} {\bibinfo {author} {\bibfnamefont {Greg}\ \bibnamefont
  {Egan}},\ }\href {http://www.gregegan.net/PERMUTATION/Permutation.html}
  {\emph {\bibinfo {title} {Permutation City}}}\ (\bibinfo {year} {1994})\
  \bibinfo {note} {accessed January 4, 2017}\BibitemShut {NoStop}%
\bibitem [{\citenamefont {Hornik}\ \emph {et~al.}(1989)\citenamefont {Hornik},
  \citenamefont {Stinchcombe},\ and\ \citenamefont {White}}]{Hornik1989359}%
  \BibitemOpen
  \bibfield  {author} {\bibinfo {author} {\bibfnamefont {Kurt}\ \bibnamefont
  {Hornik}}, \bibinfo {author} {\bibfnamefont {Maxwell}\ \bibnamefont
  {Stinchcombe}}, \ and\ \bibinfo {author} {\bibfnamefont {Halbert}\
  \bibnamefont {White}},\ }\bibfield  {title} {\enquote {\bibinfo {title}
  {Multilayer feedforward networks are universal approximators},}\ }\href
  {\doibase https://doi.org/10.1016/0893-6080(89)90020-8} {\bibfield  {journal}
  {\bibinfo  {journal} {Neural Networks}\ }\textbf {\bibinfo {volume} {2}},\
  \bibinfo {pages} {359--366} (\bibinfo {year} {1989})}\BibitemShut {NoStop}%
\bibitem [{\citenamefont {Hornik}\ \emph {et~al.}(1990)\citenamefont {Hornik},
  \citenamefont {Stinchcombe},\ and\ \citenamefont
  {White}}]{hornik:Hornik+Stinchcombe+White:1990}%
  \BibitemOpen
  \bibfield  {author} {\bibinfo {author} {\bibfnamefont {Kurt}\ \bibnamefont
  {Hornik}}, \bibinfo {author} {\bibfnamefont {Maxwell}\ \bibnamefont
  {Stinchcombe}}, \ and\ \bibinfo {author} {\bibfnamefont {Halbert}\
  \bibnamefont {White}},\ }\bibfield  {title} {\enquote {\bibinfo {title}
  {Universal approximation of an unknown function and its derivatives using
  multilayer feedforward networks},}\ }\href {\doibase
  10.1016/0893-6080(90)90005-6} {\bibfield  {journal} {\bibinfo  {journal}
  {Neural Networks}\ }\textbf {\bibinfo {volume} {3}},\ \bibinfo {pages}
  {551--560} (\bibinfo {year} {1990})}\BibitemShut {NoStop}%
\bibitem [{\citenamefont {Hornik}(1991)}]{Hornik1991251}%
  \BibitemOpen
  \bibfield  {author} {\bibinfo {author} {\bibfnamefont {Kurt}\ \bibnamefont
  {Hornik}},\ }\bibfield  {title} {\enquote {\bibinfo {title} {Approximation
  capabilities of multilayer feedforward networks},}\ }\href {\doibase
  10.1016/0893-6080(91)90009-T} {\bibfield  {journal} {\bibinfo  {journal}
  {Neural Networks}\ }\textbf {\bibinfo {volume} {4}},\ \bibinfo {pages}
  {251--257} (\bibinfo {year} {1991})}\BibitemShut {NoStop}%
\bibitem [{\citenamefont {Bell}(1990)}]{bell-a}%
  \BibitemOpen
  \bibfield  {author} {\bibinfo {author} {\bibfnamefont {John~Stuard}\
  \bibnamefont {Bell}},\ }\bibfield  {title} {\enquote {\bibinfo {title}
  {Against `measurement'},}\ }\href {\doibase 10.1088/2058-7058/3/8/26}
  {\bibfield  {journal} {\bibinfo  {journal} {Physics World}\ }\textbf
  {\bibinfo {volume} {3}},\ \bibinfo {pages} {33--41} (\bibinfo {year}
  {1990})}\BibitemShut {NoStop}%
\bibitem [{\citenamefont {Chaitin}(1987)}]{chaitin-bb}%
  \BibitemOpen
  \bibfield  {author} {\bibinfo {author} {\bibfnamefont {Gregory~J.}\
  \bibnamefont {Chaitin}},\ }\bibfield  {title} {\enquote {\bibinfo {title}
  {Computing the busy beaver function},}\ }in\ \href {\doibase
  10.1007/978-1-4612-4808-8\_28} {\emph {\bibinfo {booktitle} {Open Problems in
  Communication and Computation}}},\ \bibinfo {editor} {edited by\ \bibinfo
  {editor} {\bibfnamefont {Thomas~M.}\ \bibnamefont {Cover}}\ and\ \bibinfo
  {editor} {\bibfnamefont {B.}~\bibnamefont {Gopinath}}}\ (\bibinfo
  {publisher} {Springer},\ \bibinfo {address} {New York},\ \bibinfo {year}
  {1987})\ p.\ \bibinfo {pages} {108}\BibitemShut {NoStop}%
\bibitem [{\citenamefont {Calude}\ \emph {et~al.}(2002)\citenamefont {Calude},
  \citenamefont {Dinneen},\ and\ \citenamefont {Shu}}]{2002-glimpseofran}%
  \BibitemOpen
  \bibfield  {author} {\bibinfo {author} {\bibfnamefont {Cristian~S.}\
  \bibnamefont {Calude}}, \bibinfo {author} {\bibfnamefont {Michael~J.}\
  \bibnamefont {Dinneen}}, \ and\ \bibinfo {author} {\bibfnamefont {Chi-Kou}\
  \bibnamefont {Shu}},\ }\bibfield  {title} {\enquote {\bibinfo {title}
  {Computing a glimpse of randomness},}\ }\href {\doibase
  10.1080/10586458.2002.10504481} {\bibfield  {journal} {\bibinfo  {journal}
  {Experimental Mathematics}\ }\textbf {\bibinfo {volume} {11}},\ \bibinfo
  {pages} {361--370} (\bibinfo {year} {2002})},\ \Eprint
  {http://arxiv.org/abs/arXiv:nlin/0112022} {arXiv:nlin/0112022} \BibitemShut
  {NoStop}%
\bibitem [{\citenamefont {Calude}\ and\ \citenamefont
  {Dinneen}(2007)}]{calude-dinneen06}%
  \BibitemOpen
  \bibfield  {author} {\bibinfo {author} {\bibfnamefont {Cristian~S.}\
  \bibnamefont {Calude}}\ and\ \bibinfo {author} {\bibfnamefont {Michael~J.}\
  \bibnamefont {Dinneen}},\ }\bibfield  {title} {\enquote {\bibinfo {title}
  {Exact approximations of omega numbers},}\ }\href {\doibase
  10.1142/S0218127407018130} {\bibfield  {journal} {\bibinfo  {journal}
  {International Journal of Bifurcation and Chaos}\ }\textbf {\bibinfo {volume}
  {17}},\ \bibinfo {pages} {1937--1954} (\bibinfo {year} {2007})},\ \bibinfo
  {note} {{CDMTCS} report series 293}\BibitemShut {NoStop}%
\bibitem [{\citenamefont {Wang}(1991)}]{Wang91}%
  \BibitemOpen
  \bibfield  {author} {\bibinfo {author} {\bibfnamefont {Qui~Dong}\
  \bibnamefont {Wang}},\ }\bibfield  {title} {\enquote {\bibinfo {title} {The
  global solution of the $n$-body problem},}\ }\href {\doibase
  10.1007/BF00048987} {\bibfield  {journal} {\bibinfo  {journal} {Celestial
  Mechanics}\ }\textbf {\bibinfo {volume} {50}},\ \bibinfo {pages} {73--88}
  (\bibinfo {year} {1991})}\BibitemShut {NoStop}%
\bibitem [{\citenamefont {Diacu}(1996)}]{Diacu96}%
  \BibitemOpen
  \bibfield  {author} {\bibinfo {author} {\bibfnamefont {Florin}\ \bibnamefont
  {Diacu}},\ }\bibfield  {title} {\enquote {\bibinfo {title} {The solution of
  the $n$-body problem},}\ }\href {\doibase 10.1007/bf03024313} {\bibfield
  {journal} {\bibinfo  {journal} {The Mathematical Intelligencer}\ }\textbf
  {\bibinfo {volume} {18}},\ \bibinfo {pages} {66--70} (\bibinfo {year}
  {1996})}\BibitemShut {NoStop}%
\bibitem [{\citenamefont {Svozil}(2007)}]{svozil-2007-cestial}%
  \BibitemOpen
  \bibfield  {author} {\bibinfo {author} {\bibfnamefont {Karl}\ \bibnamefont
  {Svozil}},\ }\bibfield  {title} {\enquote {\bibinfo {title} {Omega and the
  time evolution of the $n$-body problem},}\ }in\ \href {\doibase
  10.1142/9789812770837\_0013} {\emph {\bibinfo {booktitle} {Randomness and
  Complexity, from {L}eibniz to {C}haitin}}},\ \bibinfo {editor} {edited by\
  \bibinfo {editor} {\bibfnamefont {Cristian~S.}\ \bibnamefont {Calude}}}\
  (\bibinfo  {publisher} {World Scientific},\ \bibinfo {address} {Singapore},\
  \bibinfo {year} {2007})\ pp.\ \bibinfo {pages} {231--236},\ \Eprint
  {http://arxiv.org/abs/arXiv:physics/0703031} {arXiv:physics/0703031}
  \BibitemShut {NoStop}%
\bibitem [{\citenamefont {Nietzsche}(1886, 2009-)}]{Nietzsche-JGB}%
  \BibitemOpen
  \bibfield  {author} {\bibinfo {author} {\bibfnamefont {Friedrich}\
  \bibnamefont {Nietzsche}},\ }\href
  {http://www.nietzschesource.org/\#eKGWB/JGB} {\emph {\bibinfo {title}
  {{J}enseits von {G}ut und {B}\"ose (Beyond Good and Evil)}}}\ (\bibinfo
  {year} {1886, 2009-})\ \bibinfo {note} {digital critical edition of the
  complete works and letters, based on the critical text by G. Colli and M.
  Montinari, Berlin/New York, de Gruyter 1967-, edited by Paolo
  D'Iorio}\BibitemShut {NoStop}%
\bibitem [{\citenamefont {Nietzsche}(1887, 2009-)}]{Nietzsche-GM}%
  \BibitemOpen
  \bibfield  {author} {\bibinfo {author} {\bibfnamefont {Friedrich}\
  \bibnamefont {Nietzsche}},\ }\href
  {http://www.nietzschesource.org/\#eKGWB/GM} {\emph {\bibinfo {title} {{Z}ur
  {G}enealogie der {M}oral (On the Genealogy of Morality)}}}\ (\bibinfo {year}
  {1887, 2009-})\ \bibinfo {note} {digital critical edition of the complete
  works and letters, based on the critical text by G. Colli and M. Montinari,
  Berlin/New York, de Gruyter 1967-, edited by Paolo D'Iorio}\BibitemShut
  {NoStop}%
\bibitem [{\citenamefont {Berkeley}(1710)}]{berkeley}%
  \BibitemOpen
  \bibfield  {author} {\bibinfo {author} {\bibfnamefont {George}\ \bibnamefont
  {Berkeley}},\ }\href {http://www.gutenberg.org/etext/4723} {\emph {\bibinfo
  {title} {A Treatise Concerning the Principles of Human Knowledge}}}\
  (\bibinfo {year} {1710})\BibitemShut {NoStop}%
\bibitem [{\citenamefont {Stace}(1949)}]{stace1}%
  \BibitemOpen
  \bibfield  {author} {\bibinfo {author} {\bibfnamefont {Walter~Terence}\
  \bibnamefont {Stace}},\ }\bibfield  {title} {\enquote {\bibinfo {title} {The
  refutation of realism},}\ }in\ \href {\doibase 10.1093/mind/xliii.170.145}
  {\emph {\bibinfo {booktitle} {Readings in Philosophical Analysis}}},\
  \bibinfo {editor} {edited by\ \bibinfo {editor} {\bibfnamefont {Herbert}\
  \bibnamefont {Feigl}}\ and\ \bibinfo {editor} {\bibfnamefont {Wilfrid}\
  \bibnamefont {Sellars}}}\ (\bibinfo  {publisher} {Appleton-Century-Crofts},\
  \bibinfo {address} {New York},\ \bibinfo {year} {1949})\ pp.\ \bibinfo
  {pages} {364--372},\ \bibinfo {note} {previously published in {\em Mind} {\bf
  53}, 349-353 (1934)}\BibitemShut {NoStop}%
\bibitem [{\citenamefont {Goldschmidt}\ and\ \citenamefont {Pearce}(2017,
  2018)}]{Goldschmidt2017-idealism}%
  \BibitemOpen
  \bibfield  {author} {\bibinfo {author} {\bibfnamefont {Tyron}\ \bibnamefont
  {Goldschmidt}}\ and\ \bibinfo {author} {\bibfnamefont {Kenneth~L.}\
  \bibnamefont {Pearce}},\ }\href {\doibase 10.1093/oso/9780198746973.001.0001}
  {\emph {\bibinfo {title} {Idealism: New Essays in Metaphysics}}}\ (\bibinfo
  {publisher} {Oxford University Press},\ \bibinfo {address} {Oxford, UK},\
  \bibinfo {year} {2017, 2018})\BibitemShut {NoStop}%
\bibitem [{\citenamefont {Calude}\ and\ \citenamefont
  {Longo}(2016)}]{Calude2016}%
  \BibitemOpen
  \bibfield  {author} {\bibinfo {author} {\bibfnamefont {Cristian~S.}\
  \bibnamefont {Calude}}\ and\ \bibinfo {author} {\bibfnamefont {Giuseppe}\
  \bibnamefont {Longo}},\ }\bibfield  {title} {\enquote {\bibinfo {title} {The
  deluge of spurious correlations in big data},}\ }\href {\doibase
  10.1007/s10699-016-9489-4} {\bibfield  {journal} {\bibinfo  {journal}
  {Foundations of Science}\ ,\ \bibinfo {pages} {1--18}} (\bibinfo {year}
  {2016})},\ \Eprint {http://arxiv.org/abs/CDMTCS-488} {CDMTCS-488}
  \BibitemShut {NoStop}%
\bibitem [{\citenamefont {Calude}\ and\ \citenamefont
  {Svozil}(2019)}]{svozil-2018-was}%
  \BibitemOpen
  \bibfield  {author} {\bibinfo {author} {\bibfnamefont {Cristian~S.}\
  \bibnamefont {Calude}}\ and\ \bibinfo {author} {\bibfnamefont {Karl}\
  \bibnamefont {Svozil}},\ }\bibfield  {title} {\enquote {\bibinfo {title}
  {Spurious, emergent laws in number worlds},}\ }\href {\doibase
  10.3390/philosophies4020017} {\bibfield  {journal} {\bibinfo  {journal}
  {Philosophies}\ }\textbf {\bibinfo {volume} {4}},\ \bibinfo {pages} {17}
  (\bibinfo {year} {2019})},\ \Eprint {http://arxiv.org/abs/arXiv:1812.04416}
  {arXiv:1812.04416} \BibitemShut {NoStop}%
\bibitem [{\citenamefont {Kuhn}(1970)}]{kuhn}%
  \BibitemOpen
  \bibfield  {author} {\bibinfo {author} {\bibfnamefont {T.~S.}\ \bibnamefont
  {Kuhn}},\ }\href@noop {} {\emph {\bibinfo {title} {The Structure of
  Scientific Revolutions}}},\ \bibinfo {edition} {2nd}\ ed.\ (\bibinfo
  {publisher} {Princeton University Press},\ \bibinfo {address} {Princeton,
  NJ},\ \bibinfo {year} {1970})\BibitemShut {NoStop}%
\bibitem [{\citenamefont {Feyerabend}(1974)}]{feyerabend}%
  \BibitemOpen
  \bibfield  {author} {\bibinfo {author} {\bibfnamefont {Paul~K.}\ \bibnamefont
  {Feyerabend}},\ }\href@noop {} {\emph {\bibinfo {title} {Against Method}}}\
  (\bibinfo  {publisher} {New Left Books},\ \bibinfo {address} {London},\
  \bibinfo {year} {1974})\BibitemShut {NoStop}%
\bibitem [{\citenamefont {Veyne}(1988)}]{Veyne-Greek-Mythology}%
  \BibitemOpen
  \bibfield  {author} {\bibinfo {author} {\bibfnamefont {Paul}\ \bibnamefont
  {Veyne}},\ }\href@noop {} {\emph {\bibinfo {title} {Did the {G}reeks Believe
  in Their Myths? {A}n Essay on the Constitutive Imagination}}}\ (\bibinfo
  {publisher} {University Of Chicago Press},\ \bibinfo {address} {Chicago},\
  \bibinfo {year} {1988})\BibitemShut {NoStop}%
\bibitem [{\citenamefont {Fuchs}\ and\ \citenamefont
  {Peres}(2000)}]{fuchs-peres}%
  \BibitemOpen
  \bibfield  {author} {\bibinfo {author} {\bibfnamefont {Christopher~A.}\
  \bibnamefont {Fuchs}}\ and\ \bibinfo {author} {\bibfnamefont {Asher}\
  \bibnamefont {Peres}},\ }\bibfield  {title} {\enquote {\bibinfo {title}
  {Quantum theory needs no `interpretation'},}\ }\href {\doibase
  10.1063/1.883004} {\bibfield  {journal} {\bibinfo  {journal} {Physics Today}\
  }\textbf {\bibinfo {volume} {53}},\ \bibinfo {pages} {70--71} (\bibinfo
  {year} {2000})},\ \bibinfo {note} {further discussions of and reactions to
  the article can be found in the September issue of Physics Today, {\it 53},
  11-14 (2000)}\BibitemShut {NoStop}%
\bibitem [{\citenamefont {Feynman}(1965)}]{feynman-law}%
  \BibitemOpen
  \bibfield  {author} {\bibinfo {author} {\bibfnamefont {Richard~Phillips}\
  \bibnamefont {Feynman}},\ }\href@noop {} {\emph {\bibinfo {title} {The
  Character of Physical Law}}}\ (\bibinfo  {publisher} {MIT Press},\ \bibinfo
  {address} {Cambridge, MA},\ \bibinfo {year} {1965})\BibitemShut {NoStop}%
\bibitem [{\citenamefont {Englert}(2013)}]{Englert2013}%
  \BibitemOpen
  \bibfield  {author} {\bibinfo {author} {\bibfnamefont {Berthold-Georg}\
  \bibnamefont {Englert}},\ }\bibfield  {title} {\enquote {\bibinfo {title} {On
  quantum theory},}\ }\href {\doibase 10.1140/epjd/e2013-40486-5} {\bibfield
  {journal} {\bibinfo  {journal} {The European Physical Journal D}\ }\textbf
  {\bibinfo {volume} {67}},\ \bibinfo {pages} {1--16} (\bibinfo {year}
  {2013})},\ \Eprint {http://arxiv.org/abs/arXiv:1308.5290} {arXiv:1308.5290}
  \BibitemShut {NoStop}%
\bibitem [{\citenamefont {Wolpert}(1996)}]{Wolpert:1996:LPD:1362127.1362128}%
  \BibitemOpen
  \bibfield  {author} {\bibinfo {author} {\bibfnamefont {David~H.}\
  \bibnamefont {Wolpert}},\ }\bibfield  {title} {\enquote {\bibinfo {title}
  {The lack of a priori distinctions between learning algorithms},}\ }\href
  {\doibase 10.1162/neco.1996.8.7.1341} {\bibfield  {journal} {\bibinfo
  {journal} {Neural Computation}\ }\textbf {\bibinfo {volume} {8}},\ \bibinfo
  {pages} {1341--1390} (\bibinfo {year} {1996})}\BibitemShut {NoStop}%
\bibitem [{\citenamefont {Wolpert}\ and\ \citenamefont
  {Macready}(1997)}]{Wolpert:1997:NFL:2221336.2221408}%
  \BibitemOpen
  \bibfield  {author} {\bibinfo {author} {\bibfnamefont {David~H.}\
  \bibnamefont {Wolpert}}\ and\ \bibinfo {author} {\bibfnamefont {William~G.}\
  \bibnamefont {Macready}},\ }\bibfield  {title} {\enquote {\bibinfo {title}
  {No free lunch theorems for optimization},}\ }\href {\doibase
  10.1109/4235.585893} {\bibfield  {journal} {\bibinfo  {journal} {{IEEE}
  Transactions on Evolutionary Computation}\ }\textbf {\bibinfo {volume} {1}},\
  \bibinfo {pages} {67--82} (\bibinfo {year} {1997})}\BibitemShut {NoStop}%
\end{thebibliography}


\end{document}